\documentclass[10pt]{article}
\usepackage[dvips]{graphicx}
\title{Residual Strain Dependence on Matrix Structure in RHQ-Nb$_3$Al Wires by Neutron Diffraction Measurement}
\author{Xinzhe Jin$^1$, Tatsushi Nakamoto$^1$, Takayoshi Ito$^{2\dagger}$, Stefanus Harjo$^2$, \\
Akihiro Kikuchi$^3$, Takao Takeuchi$^3$, Kiyosumi Tsuchiya$^1$, Tsutomu Hemmi$^4$, \\
Toru Ogitsu$^1$, Akira Yamamoto$^1$\\
$^1$High Energy Accelerator Research Organization (KEK), Oho, Tsukuba-shi, Ibaraki 305-0801 Japan \\
$^2$Japan Atomic Energy Agency, Tokai-mura, Naka-gun, Ibaraki 319-1195, Japan \\
$^3$National Institute for Materials Science, Sengen, Tsukuba-shi, Ibaraki 305-0047, Japan \\
$^4$Japan Atomic Energy Agency, Naka-shi, Ibaraki 311-0193, Japan \\ \\
shintetsu.kin@kek.jp}

\begin{document}
\maketitle

\begin{abstract}
We prepared three types of non-Cu RHQ-Nb$_3$Al wire samples with different matrix structures: an all-Ta matrix, a composite matrix of Nb and Ta with a Ta inter filament, and an all-Nb matrix. 
Neutron diffraction patterns of the wire samples were measured at room temperature in J-PARC ``TAKUMI.'' 
To obtain residual strains of materials, we estimated lattice constant $a$ by multi-peak analysis in the wire. 
Powder sample of each wire was measured, where the powder was considered to be strain-free. 
The grain size of all the powder samples was below 0.02 mm. 
For wire sample with the all-Nb matrix, we also obtained lattice spacing $d$ by a single-peak analysis. 
Residual strains of Nb$_3$Al filament were estimated from the two analysis results and were compared. 
Result, residual strains obtained from the multi-peak analysis showed a good accuracy with small standard deviation. 
The multi-peak analysis results for the residual strains of Nb$_3$Al filament in the three samples were all tensile residual strain in the axial direction, they are 0.12\%, 0.12\%, and 0.05\% for the all-Ta matrix, the composite matrix, and the all-Nb matrix, respectively. 
Difference in the residual strain of Nb$_3$Al filament between the composite and all-Nb matrix samples indicates that type of inter-filament materials show a great effect on the residual strain. 
In this paper, we report the method of measurement, method of analysis, and results for residual strain in the tree types of non-Cu RHO-Nb$_3$Al wires. 

\end{abstract}
\maketitle
\twocolumn
\section{Introduction}
Nb$_3$Al has advantages of better tolerance to strain/stress for stoichiometric composition over Nb$_3$Sn. 
The rapid-heating, quenching and transformation annealing (RHQ) process enables to form a stoichiometric Nb$_3$Al with fine grain structures via metastable bcc supersaturated-solid-solution. 
As a result a large critical current density of RHQ-Nb$_3A$l is achieved over the whole range of magnetic fields. 
The RHQ-Nb$_3$Al conductor is very promising for high-field applications such as particle accelerator, fusion reactor, and NMR. 
Development of the RHQ-Nb$_3$Al wires has been reported by Takeuchi $et$ $al.$[1-5]. 
The critical current $I_{\rm c}$ of A15-type superconducting wires such as Nb$_3$Al and Nb$_3$Sn dependent on the tensile strain. 
Banno $et$ $al.$ reported that $I_{\rm c}$ of RHQ-Nb$_3$Al wires decreases with an increase in the tensile strain\cite{Banno}. 

Since RHQ-Nb$_3$Al wires are composites, thermal phase stress during sample preparation (heat treatment) can be generated due to difference in the coefficient of thermal expansion (CTE) among constituent phases. 
The thermal stresses are also different for samples with different matrix.
At the room temperature, the Nb$_3$Al filament has residual strain induced by the difference in the thermal contraction between the filament and the matrix material during cool-down from the heat treatment to the room temperature. 
It is important to study the residual strain of RHQ-Nb$_3$Al to thoroughly understand its $J_{\rm c}$ dependence on the strain, but it has not been studied yet. 
The CTE of Nb$_3$Al is smaller than that of Cu and larger than that of Ta or Nb\cite{Murase,Seeber}. 
Therefore, the residual strain of Nb$_3$Al filament is determined by the complicated interaction with the compressive and the tensile stresses by the Cu stabilizer and the Ta or Nb matrix. 
To understand the mechanism of generation of the residual strain in RHQ-Nb$_3$Al wires, it is necessary to investigate the effect of each component. 
In this study, we focused on the dependence of this residual strain on the matrix structure at room temperature. 
For this purpose, the Cu plating was removed after heat treatment for Nb$_3$Al phase transformation. 

Residual strain of Nb$_3$Sn in bronze processed Nb$_3$Sn wire has been estimated to be about -0.2 -- -0.1\%\ at the room temperature by a single-peak analysis in the axial direction\cite{Oguro, Awaji}. 
The residual strain of Nb$_3$Sn was shown to have a significant effect by Cu. 
The CTE of Nb$_3$Sn is smaller than that of Cu by 9.6 $\times$ 10$^{-6}$ K$^{-1}$. 
This difference is about four times larger than the difference between the CTEs of Nb$_3$Al and Ta (or Nb) in non-Cu Nb$_3$Al wires. 
Considering only the magnitude of strain in the axial direction, the residual strain of Nb$_3$Al filament in the RHQ-Nb$_3$Al wires was expected to be smaller than that of Nb$_3$Sn in the bronze processed Nb$_3$Sn wires. 
Thus, high precision is necessary for analyzing the residual strain in RHQ-Nb$_3$Al wires. 

\section{Experimental}
\subsection{Sample preparation}

\begin{figure}[t]
\centering
\includegraphics[width=3.2in]{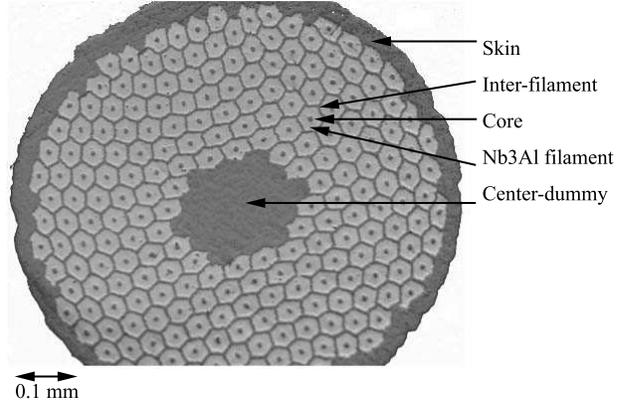}
\caption{Cross section of sample A. The skin, inter-filaments, core, and center-dummy are all made of Ta.}
\label{crosssection}
\end{figure}

We prepared three types of the RHQ-Nb$_3$Al wire samples without Cu stabilizer and with different matrix materials  (samples A, B, and C). Main parameters are summarized as shown in Table \ref{t_m}. 
The cross section of sample A is shown in Fig. \ref{crosssection}. 

The following is the sample preparation process. 
Firstly, a mono filament was prepared by the jelly-roll method\cite{Takeuchi2}. 
Nb and Al sheets with a stoichiometric composition were rolled onto a pure Nb or a pure Ta core. 
The mono filament was bundled into a Cu conduit with Ta or Nb matrix, and a multi filament was fabricated by the wire drawing process. 
After drawing the wires, the Cu conduit was removed, and they were heated to above 2000 $^\circ $C by RHQ treatment. 
Then, the Cu stabilizer was plated. 
The cross-sectional diameter of the drawn wires was 1 mm, and they were heated at 800 $^\circ $C for 10 h. 
Finally, the Cu plating was removed. 

To determine the residual strain of the Nb$_3$Al filament in the composite, a sample representing a strain-free state is necessary. 
Since removal of the Ta or Nb matrix surrounding the Nb$_3$Al filament is very difficult, the samples A, B and C were powderized. 
The powder samples were prepared with a particle size below 0.02 mm to achieve a strain-free state. 

\begin{table}[t]
\caption{Specification of samples A, B, and C.}
\label{t_m}
\centering
\begin{tabular}{ccccc}
\hline
Sample & A & B & C \\
\hline
Skin & Ta & Nb & Nb \\
Inter-filament & Ta & Ta & Nb \\
Core & Ta & Nb & Nb \\
Center-dummy & Ta & Nb & Nb \\
Matrix/Nb$_3$Al ratio & 0.8 & 0.8 & 0.8 \\
Number of Nb$_3Al$ filament & 222 & 222 & 144 \\
Diameter of Nb$_3Al$ filament (mm) & 0.04 & 0.04  & 0.05 \\
Outer diameter (mm) & 0.7 & 0.7 & 0.7 \\
\hline
\end{tabular}
\end{table}

\begin{figure}[t]
\centering
\includegraphics[width=3.5in]{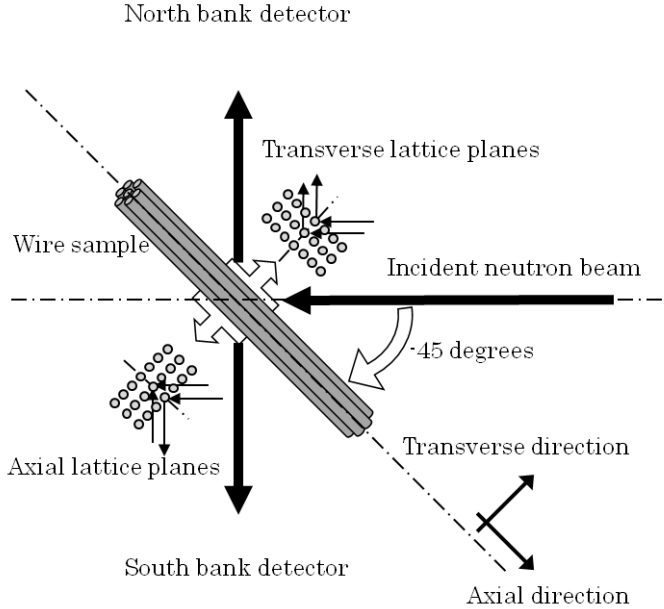}
\caption{2D diagram of neutron measurement of wire sample along transverse and axial directions with two detectors in J-PARC TAKUMI.}
\label{diffraction}
\end{figure}

\subsection{Sample measurements}

The  ``TAKUMI'', a time-of-flight (TOF) neutron diffractometer at MLF/J-PARC, was chosen to conduct this study\cite{Harjo1,Harjo2}. 
Multiple peaks in the diffraction pattern were measured simultaneously using two fixed detectors placed at two opposite banks whose diffraction angles with respect to the incident beam were fixed to be -90$^\circ$ and 90$^\circ$.
Therefore, the diffraction patterns for the wire in the transverse and axial lattice directions can be obtained simultaneously for a beam incident angle of 45$^\circ$, as shown in Fig. \ref{diffraction}. 

For neutron diffraction measurement, seven wires of 20 mm length were bundled. 
The measurements were performed at room temperature, and time for a single measurement was 20 minutes. 
To confirm the efficacy of multi-peak analysis by comparing with the single-peak analysis, the sample C was measured four times under the same measurement conditions. 
There may be slight differences in position among the four measurements because multiple samples were set on the goniometer and they were scanned through the measurement. 
Measurement for the powder samples in sealed cylindrical aluminum holders were carried out by using a radial collimator of 2 mm width\cite{Torii}. 
The bore and outer diameters of all the aluminum holders are 8 and 10 mm, respectively. 
Calibration of TOF and lattice spacing $d$ was performed using CeO$_2$ powder as a standard sample certificated by NIST, and conversion parameters were determined. 
Because the volume fraction of Nb$_3$Al was small in all the samples, we used the high intensity mode with a sample-slit width of 5 mm for the incident neutron beam.. 
The experiments were conducted at a proton beam power of 120 kW. 

\section{Results and discussion}
\subsection{Neutron diffraction patterns}

\begin{figure}[t]
\centering
\includegraphics[width=3.2in]{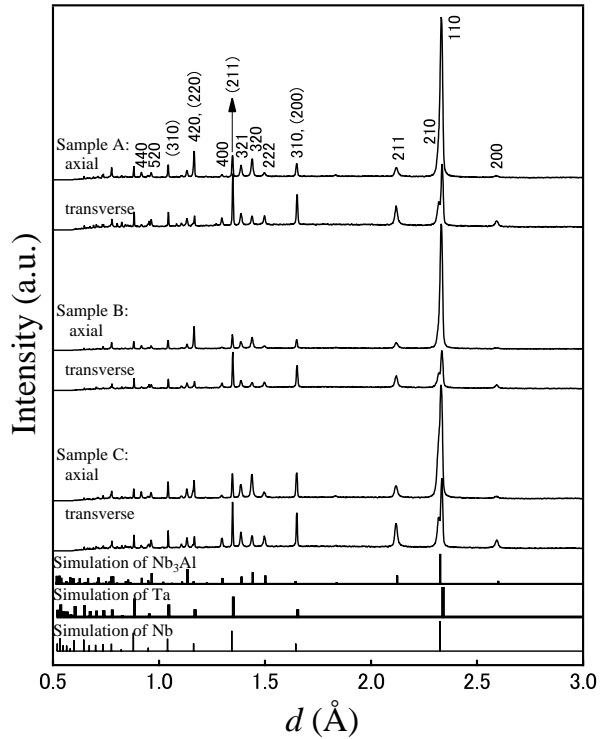}
\caption{Neutron diffraction patterns of wire samples with powder simulations of Nb$_3$Al, Ta, and Nb. 
The Miller indices of peaks for Nb$_3$Al and Ta (in parentheses) were shown in axial pattern of sample A.}
\label{ND_allwire}
\end{figure}

Neutron diffraction patterns of the wire samples are shown in Fig. \ref{ND_allwire}. 
The powder neutron diffraction simulations of Nb$_3$Al, Ta, and Nb were calculated by using Rietan 2000\cite{Izumi}. 
In all the patterns, the Nb$_3$Al and matrix material (Ta or Nb) phases were observed without any impurity phase. 
All the samples showed a difference in the relative intensity ratio of each hkl peak between the transverse and axial directions. 
This difference indicates that crystallites in the RHQ-Nb$_3$Al wires have preferred orientation. 

\begin{figure*}[t]
\begin{minipage}{0.5\hsize}
\centering
\includegraphics[width=3.2in]{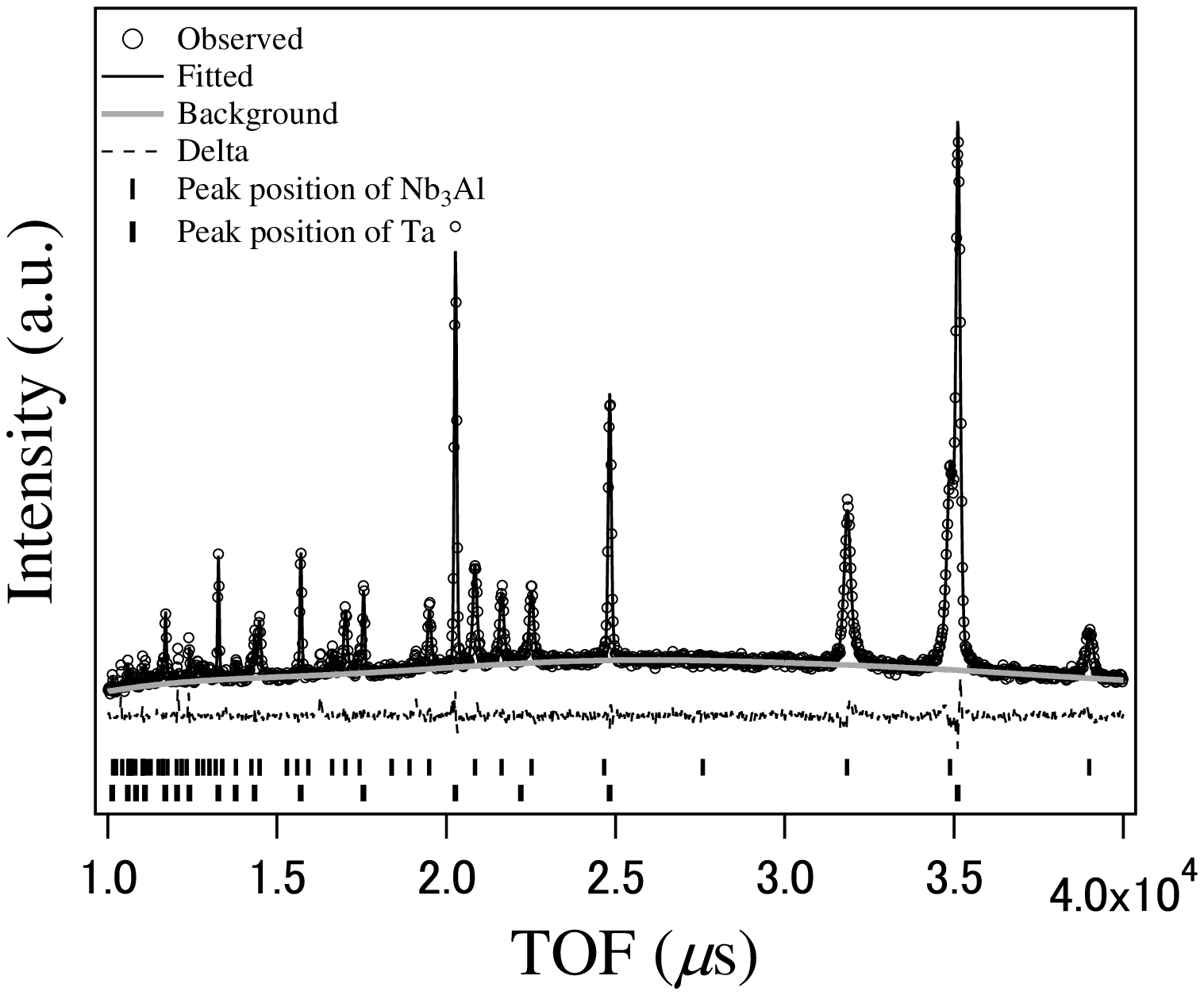}
\caption{Multi-peak analysis results for sample A in transverse direction. A delta is the difference in the intensity between the measurement and analysis results. The analyzed peak positions of Nb$_3$Al and Ta were shown with vertical lines.}
\label{ND_A_multi_t}
\end{minipage}
\begin{minipage}{0.5\hsize}
\centering
\includegraphics[width=3.2in]{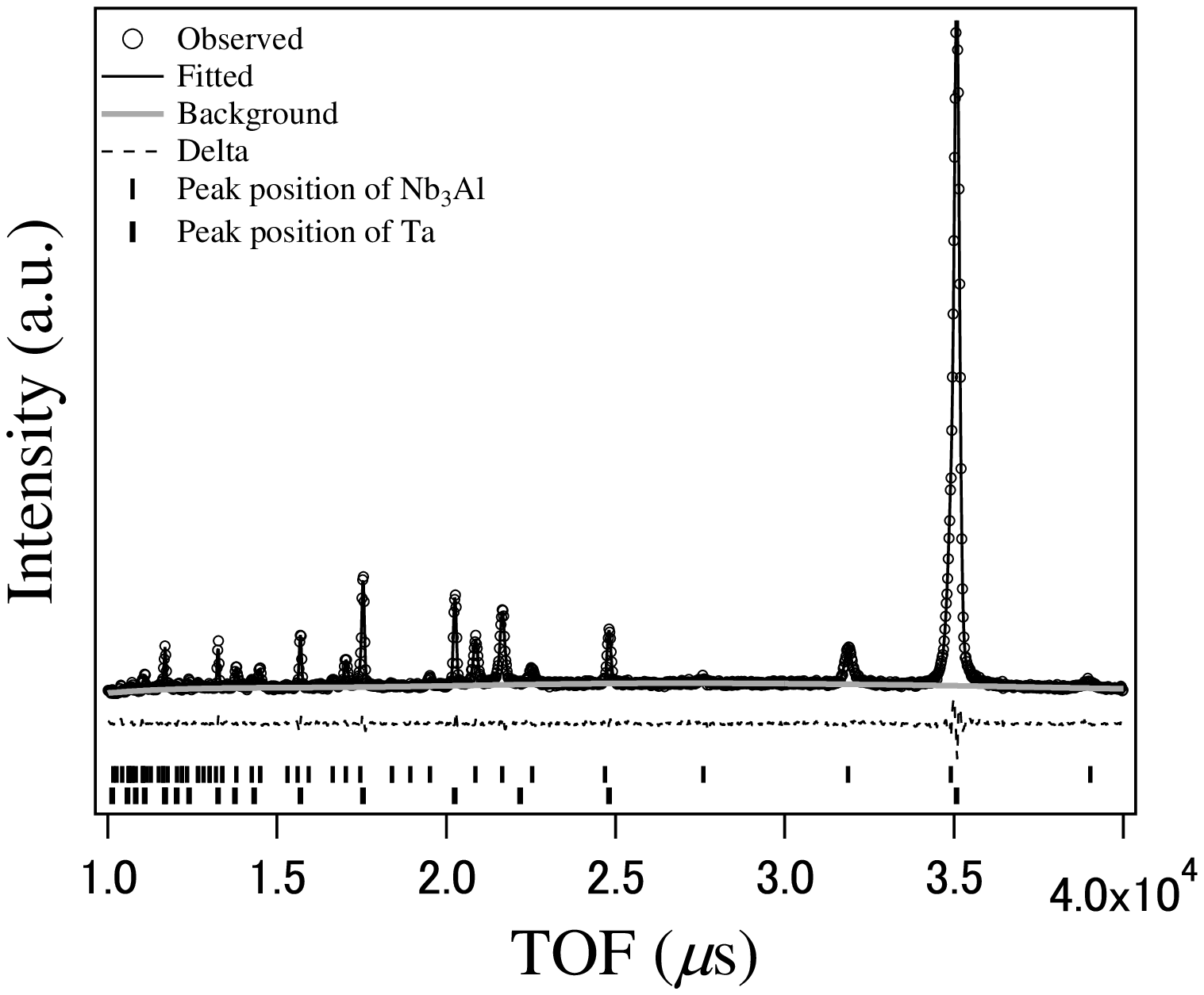}
\caption{Multi-peak analysis results for sample A in axial direction. A delta is the difference in the intensity between the measurement and analysis results. 
The analyzed peak positions of Nb$_3$Al and Ta were shown with vertical lines.}
\label{ND_A_multi_a}
\end{minipage}
\end{figure*}

\subsection{Multi-peak analysis and results}

To obtain the residual strain, we firstly determined lattice constant by a multi-peak analysis for a specific phase in the neutron diffraction pattern. 
In this study, we applied the multi-peak analysis to the composite wires.
The lattice constant $a_{\rm m}$ of the wire determined by the multi-peak analysis was fitted by using Z-Rietveld (ver. 0.9.34) for more than 10 peaks\cite{Oishi}. 
The Z-Rietveld was developed for Reitveld analysis of powder sample, it is also can fit to non-powder model by setting each peak on independence. 

From diffraction patterns in the axial and transverse directions in Fig. \ref{ND_allwire} , it is clearly shown that the wire samples have strong textures. 
Therefore, the intensity of each peak for a phase was fitted as a free fitting parameter without using random model or any preferred orientation function. 
This is different from Rietveld analysis. 

Cubic crystal structure of Nb$_3$Al can deformed to the non-cubic structures under anisotropic lattice strain in the composite wire. 
To estimate the lattice constant $a_{\rm m}$, the lattice spacings $d$ of each hkl peak were fitted by the linear least-square method that was simulated to lattice space group $P_{m\bar{3}n}$ of the powder. 
Therefore, this is statistical analysis by unification of lattice structures. 

The powder samples were also analyzed by the multi-peak analysis. 
The peak intensities also were fitted in independence of each peak. 
The lattice constant for powder sample was defined as $a_{\rm 0}$. 

As an example, the analysis results for the sample A comprised of all Ta matrix in transverse and axial directions are shown in Figs. \ref{ND_A_multi_t} and \ref{ND_A_multi_a}, respectively. 
The differences in the intensities between the measured and the fitted patterns (deltas) have small scatters, indicating a good fitting. 
The fitted peak positions of Nb$_3$Al and Ta are shown in the figures with vertical lines. 

Table \ref{t_a1} shows the lattice constants $a_{\rm m}$ and $a_{\rm 0}$ for the wire and powder samples, respectively. 
All the lattice constants $a_{\rm m}$ were larger than the corresponding lattice constant $a_{\rm 0}$. 
This indicates that the Nb$_3$Al lattice was extended along the transverse and axial directions in all the wire samples. 

The residual strain $\epsilon$ by the multi-peak analysis is given as 

\begin{eqnarray*}
\epsilon = 
\frac{a_{\rm m} - a_{\rm 0}}{a_{\rm 0}}. 
\end{eqnarray*}

\begin{table}[h]
\caption{Lattice constants {\rm $a_{\rm m}$} and {\rm $a_{\rm 0}$} of materials for three samples. 
The $a_{\rm m}$ and $a_{\rm 0}$ are lattice constants of the wire and powder samples, respectively, obtained by the multi-peak analysis.}
\centering
\label{t_a1}
\begin{tabular}{ccc|cc}
\hline
Sample&Material&Direction&$a_{\rm m}$ (\AA)& $a_{\rm 0}$ (\AA)\\
\hline
\raisebox{-4ex}[0cm][0cm]{A} & \raisebox{-2ex}[0cm][0cm]{Nb$_3$Al} & transverse & 5.1877 & \raisebox{-1.5ex}[0cm][0cm]{5.1852}\\
                                      &                                       & axial & 5.1919 & \\
                                      & \raisebox{-2ex}[0cm][0cm]{Ta} & transverse & 3.3027 &\raisebox{-1.5ex}[0cm][0cm]{3.3016}\\
                                      &                                 & axial & 3.2997 & \\   
\hline                          
\raisebox{-2ex}[0cm][0cm]{B} & \raisebox{-2ex}[0cm][0cm]{Nb$_3$Al} & transverse & 5.1879 & \raisebox{-1.5ex}[0cm][0cm]{5.1856}\\
                                      &                                       & axial & 5.1919 & \\                          
\hline
\raisebox{-4ex}[0cm][0cm]{C} & \raisebox{-2ex}[0cm][0cm]{Nb$_3$Al} & transverse & 5.1867 & \raisebox{-1.5ex}[0cm][0cm]{5.1860}\\
                                      &                                       & axial & 5.1887 & \\
                                      & \raisebox{-2ex}[0cm][0cm]{Nb} & transverse & 3.3004 &\raisebox{-1.5ex}[0cm][0cm]{3.3019}\\
                                      &                                 & axial & 3.2991 & \\
\hline
\end{tabular}
\end{table}

\begin{figure}[t]
\centering
\includegraphics[width=3.2in]{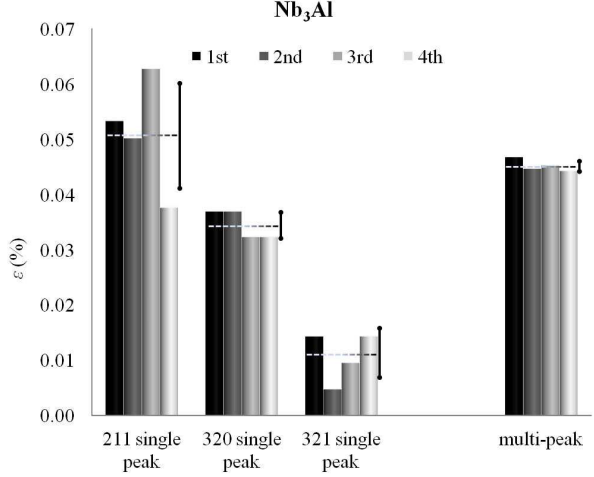}
\caption{Residual strains of Nb$_3$Al filament in sample C from single-peak and multi-peak analyses in axial direction. 
The dashed line shows the average of the four measurements. 
The circular arrow line shows error by the standard deviation of that. 
In the multi-peak analysis, strain was calculated from lattice constant $a$ that was analyzed by using Z-Rietveld.}
\label{strain}
\end{figure}

\subsection{Comparison with single-peak analysis}

In order to confirm the efficacy of multi-peak analysis, the comparison with single-peak analysis was carried out.
The lattice spacing $d$ for the 211, 320 and 321 peaks of Nb$_3$Al in the sample C and its powder was estimated in the axial direction with Gaussian fitting. 
In the single-peak analysis, we defined the lattice spacing of the wire and powder samples as $d_{\rm s}$ and $d_0$, respectively. 
By considering the powder samples to be strain-free, the residual strains $\epsilon$ can be evaluated by 

\begin{eqnarray*}
\epsilon = 
\frac{d_{\rm s} - d_{\rm 0}}{d_{\rm 0}}. 
\end{eqnarray*}

In the four measurement of the sample C, results for residual strain obtained by the single-peak and multi-peak analyses were shown in Fig. \ref{strain}. 
The dashed line shows the averaged residual strain of the four measurements for the 211 peak, 320 peak, 321 peak, and multi-peak as 0.051\%, 0.035\%, 0.011\%, and 0.045\%, respectively. 
The variations in single-peak analysis results indicate that residual strain depends on the hkl peaks, it maybe related to different diffraction elastic stiffness\cite{Ito}. 
Considering these variation in hkl peaks, residual strains analyzed from multi-peak analysis method show weighted-averaged values which may express macroscopic residual strains. 
The standard deviation of the four residual strains obtained from the multi-peak analysis is 0.001\%. 
This value is 3--9 times smaller than that obtained from single-peak analysis. 
It indicates that statistics by multi-peak analysis in wire certainly is valid and feasible to improve accuracy. 

\subsection{Residual strains in wire with different matrix}

\begin{figure*}[t]
\centering
\includegraphics[width=6.5in]{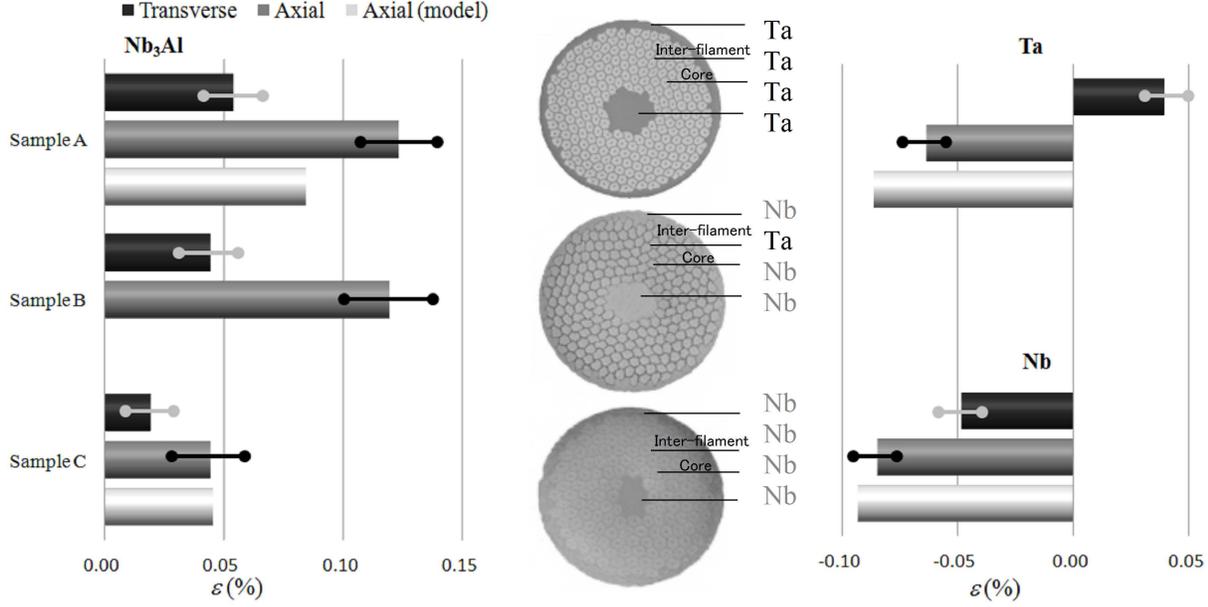}
\caption{Residual strains obtained from the multi-peak analysis. 
The residual strains in samples A and C because of thermal contraction were calculated in axial direction shown as model.}
\label{strain_cross}
\end{figure*}

In the samples A, B, and C, the residual strains of the Nb$_3$Al filament obtained from the multi-peak analysis are shown in Fig. \ref{strain_cross} with the corresponding cross section of wire. 
The values of error bars were obtained by summation of systematic error 0.1\% and the error obtained from Z-Rietveld. 
All the residual strains had tensile strain at the axial direction. 
In all the samples, the residual strains of Nb$_3$Al filament in the axial direction were larger than that in the transverse direction. 
It indicates that thermal stress shows a significant effect in the axial direction. 
Similar behavior can be seen in strain distribution of the Nb3Sn\cite{Awaji}. 

In the given direction, the samples A and B showed almost the same residual strain in the Nb$_3$Al filament. 
This result indicates that the residual strain of the Nb$_3$Al filament is dominantly influenced by the adjacent inter-filament matrix of Ta and the effects of the skin, core, and center dummy are relatively small. 
The residual strain of the Nb$_3$Al filament in the sample B was larger than that in the sample C by about 0.08 \%\ and 0.02 \%\ in the axial and transverse directions, respectively. 
The residual strain of Nb$_3$Al filament has a great dependence on the materials of inter-filament. 

We also estimated the residual strain of the matrixes of  Ta and Nb by the multi-peak analysis. 
The crystal structures of Ta and Nb have the same space group $P_{m\bar{3}n}$ and almost same parameters. 
Thus, it is difficult to distinguish Ta and Nb peaks in the diffraction patterns of the sample B in which both Ta and Nb are utilized for the matrix materials. 
The analysis results for the matrix materials in the samples A and C are shown in Fig. \ref {strain_cross}. 
In the transverse direction, the Ta and Nb show tensile and compressive residual strain, respectively. 
It indicates the samples A and C have different generation mechanisms of residual strain in the transverse direction. 

\subsection{Model calculation of residual strain induced by thermal contraction }

We calculated residual strains of Nb$_3$Al filament and matrix materials by a thermal contraction model for samples A and C in the axial direction. 
In this model, the transverse strains were assumed to have no effect on the axial strains. 
During heat treatment, as the temperature decreased below 800 $^\circ$C, axial residual strain was assumed to be generated 550 $^\circ$C ($T_{\rm H}$)\cite{Ito, Osamura}. 
Then, as temperature decreased to room temperature ($T_{\rm R}$), the residual strain increased with an increase in the stress between the Nb$_3$Al filament and the matrix material owing to the difference in their CTE. 

\begin{figure}[t]
\centering
\includegraphics[width=3.2in]{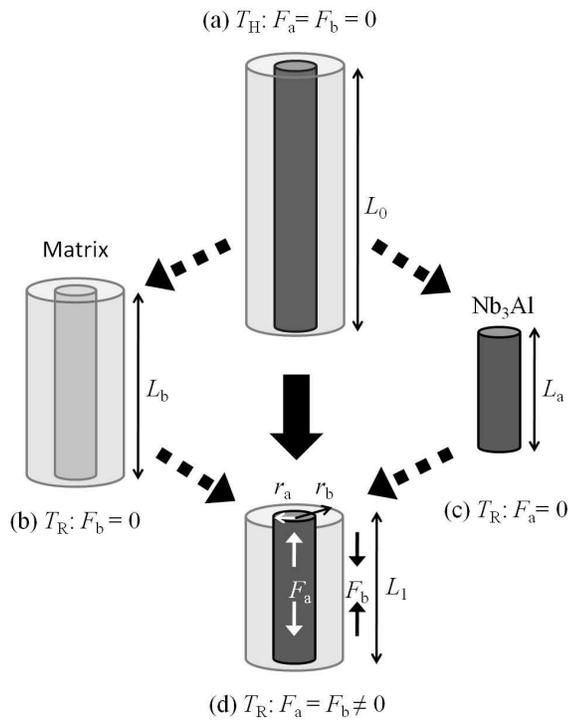}
\caption{Model representation of thermal strain in axial direction. 
The residual strain depends on the difference in CTE, Young's modulus, and cross-sectional area between Nb$_3$Al and the matrix material.}
\label{model}
\end{figure}

CTE of Nb$_3$Al is larger than that of Ta or Nb; therefore, the Nb$_3$Al filament has tensile residual strain in the axial direction. 
The model representation of thermal contraction is shown in Fig. \ref{model}. 
In Fig. \ref{model} (a), the axial length of the wire is defined as $L_0$, at the 550 $^\circ$C when the wire was free of residual strain. 
The axial lengths of the Nb$_3$Al filament and the matrix material at room temperature are $L_a$ and $L_b$, respectively, considering their thermal contraction separately as shown in Fig. \ref{model} (b) and (c). 
In actuality, the composite wire shrank to length $L_1$ with residual strains between the Nb$_3$Al filament and the matrix material at room temperature, as shown in Fig. \ref{model} (d). 
The residual strains of Nb$_3$Al filament and matrix materials in the axial direction were calculated as follows. 

At room temperature, the stresses $\sigma$ between the Nb$_3$Al filament and the matrix material in axial direction are given as 
\begin{eqnarray}
\int_{0}^{r_a} {\sigma _adr} + \int_{r_a}^{r_b} {\sigma _bdr} = 
0
\end{eqnarray} 
As $\sigma$ = $E\epsilon$, 
\begin{eqnarray}
\epsilon_{\rm a} = 
\frac{L_1 - L_a}{L_a}, and 
\end{eqnarray}
\begin{eqnarray}
\epsilon_{\rm b} = 
\frac{L_1 - L_b}{L_b},
\end{eqnarray} 
eq. (1) can be written
\begin{eqnarray}
E_{\rm a}f_{\rm a}\frac{L_1-L_a}{L_a} = 
- E_{\rm b}f_{\rm b}\frac{L_1-L_b}{L_b}, 
\end{eqnarray} 
where $E$ is the Young's modulus, $f$ is the volume fraction, and $\epsilon$ is the residual strain between the Nb$_3$Al filament and the matrix material. 
The thermal strains $\epsilon'$ are expressed as 
\begin{eqnarray}
\epsilon'_{\rm a} = 
\int_{T_H}^{T_R} {(- C_{\rm a}) dT} =
\frac{L_{\rm a} -L_0}{L_{\rm 0}}
\end{eqnarray} 
\begin{eqnarray}
\epsilon'_{\rm b} = 
\int_{T_H}^{T_R} {(- C_{\rm b}) dT} =
\frac{L_{\rm b} - L_0}{L_{\rm 0}}.
\end{eqnarray} 
From eqs. (2), (3), (5), and (6), the residual strains can be written as 
\begin{eqnarray}
\epsilon_{\rm a} = 
\frac{L_1}{L_0(1+\epsilon'_{\rm a})} - 1
\end{eqnarray} 
\begin{eqnarray}
\epsilon_{\rm b} = 
\frac{L_1}{L_0(1+\epsilon'_{\rm b})} -1 .
\end{eqnarray} 
Then, 
\begin{eqnarray}
\frac{L_1}{L_0} = 
\frac{(E_af_a + E_bf_b)(1+\epsilon'_{\rm a})(1+\epsilon'_{\rm b})}{E_{\rm a}f_{\rm a}(1 + \epsilon'_{\rm b}) + E_{\rm b}f_{\rm b}(1 + \epsilon'_{\rm a})}
\end{eqnarray}
from eqs. (4), (5), and (6). 

Eqs. (7), (8), and (9) show that the residual strains are 
\begin{eqnarray*}
\epsilon_{\rm a} = \frac{(E_af_a + E_bf_b)(1 + \epsilon'_{\rm b})}{E_{\rm a}f_{\rm a}(1 + \epsilon'_{\rm b}) + E_{\rm b}f_{\rm b}(1 + \epsilon'_{\rm a})} - 1
\end{eqnarray*} 
\begin{eqnarray*}
\epsilon_{\rm b} = \frac{(E_af_a + E_bf_b)(1 + \epsilon'_{\rm a})}{E_{\rm a}f_{\rm a}(1 + \epsilon'_{\rm b}) + E_{\rm b}f_{\rm b}(1 + \epsilon'_{\rm a})} -1.
\end{eqnarray*}

The calculation parameters for Nb$_3$Al, Ta, and Nb at room temperature are shown in Tables \ref{t_m} and \ref{t_c}. 

\begin{table}[h]
\begin{center}
\caption{CTE and Young's modulus of materials at room temperature\cite{Murase,Seeber}.}
\label{t_c}
\begin{tabular}{cccc}
\hline
\raisebox{1.5ex}[0cm][0cm]{Material}& \shortstack{CTE \\ (10$^{-6}$K$^{-1}$)} & \shortstack{Young's modulus \\ (GPa)} \\
\hline
Ta & 6.4 & 185 \\
Nb & 7.1 & 103 \\
Nb$_3$Al & 8.75 & 169 \\
Nb$_3$Sn & 7.2 & 165 \\
\hline
\end{tabular}
\end{center}
\end{table}

The calculation results for residual strain are shown in Fig. \ref{strain_cross}. 
In sample C, the calculation results for Nb$_3$Al filament and Ta, which was similar to result obtained by multi-peak analysis. 
In sample A, residual strain obtained by the model calculation was less than that obtained by the multi-peak analysis for the Nb$_3$Al filament and the former was larger than the latter for the Ta. 
This indicates that the strain mechanism for the all-Ta matrix is different from that for the all-Nb matrix in axial direction. 
Considering the different strain mechanism of samples A and C in transverse direction, it may be related to effect from the transverse direction due to the high hardness of the Ta and the complicated structure of cross section. 

\section{Conclusion}

The residual strains for Nb$_3$Al filament and matrix materials in RHQ-Nb$_3$Al wires were measured by neutron diffraction at room temperature. 
We have estimated the lattice constants $a_{\rm m}$ for three types wires by multi-peak analysis having anisotropic lattice strains in axial and transverse directions of wire, due to the different diffraction elastic stiffness in lattice planes, and the sample configuration. 
The multi-peak analysis showed good accuracy in the estimation of residual strain in RHQ-Nb$_3$Al wires. 
We have estimated the tensile residual strain of the Nb$_3$Al filament and the compressive strain of the Ta and Nb in the  axial direction, by analyzing the lattice constant $a_{\rm m}$ . 
By comparing the all-Nb matrix and the composite matrix samples, the residual strains is greatly affected by the type of inter-filament material. 
We also obtained the difference in the residual strain of Nb$_3$Al filament between the all-Ta matrix and all-Nb matrix samples in the RHQ-Nb$_3$Al wires, and the residual strains were calculated by thermal strain model in axial direction. 
Result, the all-Ta matrix and the all-Nb matrix samples showed different strain mechanisms in the axial and transverse direction. 

This study was conducted under the LHC high luminosity upgrade project.

\end{document}